\documentstyle[psfig,conf_iap]{article}

\begin{document}

\heading{The power spectrum of the Lyman--$\alpha $ clouds}

\par\medskip\noindent

\author{Luca Amendola$^{1}$, Sandra Savaglio$^{2}$}

\address{Osservatorio Astronomico di Roma, Viale del Parco Mellini 84,
I-00136 Roma, Italy}

\address{European Southern Observatory, Schwarzschild Str. 2, 
D-85748 Garching, Germany}

\begin{abstract}
We estimate the 1--D power spectrum and the integral density of neighbors in
14 QSO lines of sight. We discuss their variation with respect to
redshift and column density and compare the results with standard CDM models.
\end{abstract}

\bigskip
\medskip\noindent
We analysed 14 QSO lines of sight (Cristiani et al. 1997
and references therein), searching for clustering and
evolution. We employed two related quantities, the one--dimensional power
spectrum (PS) $P(k)$  and the integrated density of neighbors $\rho _n,$
expressed in terms of the 3--dim power spectrum $P_3(k)$:

\begin{equation}
P(k)=\frac 1{2\pi }\int_k^\infty P_3(k^{\prime })k^{\prime }dk^{\prime }
\end{equation}
\begin{equation}
\rho _n=1+\frac 1{2\pi }\int_0^\infty P(k)W(kL)dk  \label{p1rho}
\end{equation}

\noindent
where $W(x)$ is the 1--dim top--hat window function, $W(x)=\sin (x)/x$. 
The 1--dim power spectrum in units of the noise and averaged over all the
lines of sight gives no detectable sign of clustering nor of
characteristic scales as those found in the galaxy distribution. However
the 1--dim power spectrum is extremely noisy and cannot be easily compared
to theoretical predictions. To perform this comparison, we estimated
the integrated density of neighbors at various scales.  The theoretical
linear CDM--like galaxy spectrum is
\[
P_{cdm}\left( k,\Gamma ,z\right) =AkT(k,\Gamma )^2(1+z)^{-2}
\]
where $\Gamma $ is a theoretical free parameter (equal to $\Omega _0h$ in
CDM models) and $A$ is a normalization which is to be fixed by the local
galaxy clustering. We normalize as usual putting the variance in 8 Mpc $%
h^{-1}$ spherical top--hat cells equal to unity. Further, we correct the CDM
spectrum for the large-- and small--scale redshift distortion and for the IGM
Jeans length. According to the standard model of biased gravitational
instability, the power spectrum gives density fluctuations that  originates
the {\rm Ly}$\alpha $~forest. Bi \& Davidsen (1997) show that the evolution
of the standard PS gives a decrease of the HI column density of the same 
{\rm Ly}$\alpha $~cloud with decreasing redshift. 
Aim of our analysis is to determine the power spectrum of the
Lyman--$\alpha $ forest in terms both of redshift and of column density $\log
N_{HI}$. We express the Ly$\alpha $ PS as
\[
P_{{\rm Ly}\alpha }(k)=b_{{\rm Ly}\alpha }^2(z,\log N_{HI})P_{cdm}(k,\Gamma
,z)~.
\]

\begin{figure}[tbp]
\centerline{\vbox{
\psfig{figure=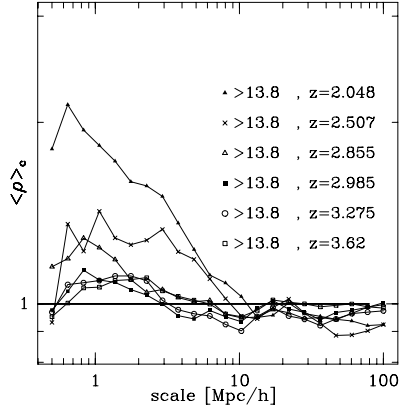,height=5.5cm,angle=0}
}}
\caption{The density of neighbors in different sets and for $\log
N_{HI}\geq 13.8$.}
\label{f5}
\end{figure}

\noindent
Clearly, $b_{{\rm Ly}\alpha }^2(z)=const$ means that the clouds evolve
linearly, and $b_{{\rm Ly}\alpha }^2(z)=1$ that they are clustered according
to CDM\ expectations. 
For the set of the seven highest redshift systems ($<z>=3.36$) 
we see that in the density of neighbors there is no detectable
clustering for any
column density cut--off. For the lowest redshift
subset, we see a clear increase in clustering with $\log N_{HI}$. Comparing
with the CDM curve, we see that only the lines with $\log N_{HI}\geq 13.8$
and $\geq 14$ are near the CDM linear predictions, while all the other sets
have lower clustering. In Fig.~\ref{f5} we show the clustering trend in six
subsets of our sample. The clustering increases regularly from $z=4$
to $z=2$, although the increase is quite faster at low $z$. 
Assuming as a reference
the CDM predictions with $\Gamma =0.2$, $\Omega ^{0.6}/b=1$, and linear
evolution, we can estimate the evolution function $b_{{\rm Ly}\alpha
}^2(z,\log N_{HI})$. As mentioned the real evolution is likely to
be both in column density and in clustering strength. If one
arbitrarily assumes that the average column density is time--independent, then
the evolution in redshift turns out to be extremely fast: a trend as $(1+z)^5$
seems to fit our results  for high column density cut--offs. One can obtain
more reasonable values, i.e. a behavior closer to the linear trend,  {\it %
only }assuming at the same time a decrease of the average column density.
The function $b_{{\rm {Ly}\alpha }}^2$  can be fitted as 
\begin{equation}
b_{{\rm {Ly}\alpha }}^2=0.46(\log N_{HI}^{})^{4.9}(1+z)^{-10.5}
\end{equation}
As an interesting consequence, we see that the clustering evolves linearly,
i.e. $b_{{\rm {Ly}\alpha }}^2$ is constant in redshift if and only if $\log
N_{HI}^{}\sim (1+z)^{10.5/4.9}\sim (1+z)^2.$

\end{document}